\documentstyle[12pt]{article}
 
\begin{document} 

\baselineskip 22pt 

\begin{center}
{\Large
\bf
Model Independent Calculation of\\
${\cal{B}}({\bar{B}}^0\rightarrow D^{(*)+} \tau^- {\bar{\nu}})/
{\cal{B}}({\bar{B}}^0\rightarrow D^{(*)+} e^- {\bar{\nu}})$}\\
\vspace{1.0cm}
Dae Sung Hwang${}^{(a)}$ and Do-Won Kim${}^{(b)}$\\
{\it{a: Department of Physics, Sejong University, Seoul 143--747,
Korea}}\\
{\it{b: Department of Physics, Kangnung National University,
Kangnung 210-702, Korea}}\\
\vspace{2.0cm}
{\bf Abstract}\\
\end{center}

\noindent
Using the formulas for the $d\Gamma / dq^2$ distribution
with non-zero lepton mass and experimentally determined form factors,
we calculate the $d\Gamma (D^{(*)+} l^- {\bar{\nu}})/ dq^2$
spectra and branching fractions for $l=e$, $\mu$ and $\tau$.
We obtain the results
${\cal{B}}({\bar{B}}^0\rightarrow D^{+} \tau^- {\bar{\nu}})/
{\cal{B}}({\bar{B}}^0\rightarrow D^{+} e^- {\bar{\nu}})=
0.278^{+0.049}_{-0.035}$
and
${\cal{B}}({\bar{B}}^0\rightarrow D^{*+} \tau^- {\bar{\nu}})/
{\cal{B}}({\bar{B}}^0\rightarrow D^{*+} e^- {\bar{\nu}})=
0.256^{+0.014}_{-0.013}$ .
Since we used the experimentally measured form factors, these results are
independent of theoretical models of form factors.
\\

\vfill


\thispagestyle{empty}
\pagebreak

\baselineskip 22pt

The heavy quark effective theory (HQET) allows the form factors of the heavy
to heavy meson transitions to be expressed in terms of the Isgur-Wise function
\cite{iw}.
The HQET can be applied to the $B$ to $D$ transitions, and the precise
measurements of the $q^2$ spectra of the exclusive semileptonic decays
${\bar{B}}^0\rightarrow D^{(*)+} l^- {\bar{\nu}}$ with electron or muon have
been used for the determination of the Isgur-Wise function \cite{cleo95, cleo97}.
However, the exclusive semileptonic decays with tau lepton have not been
measured accurately yet, and these measurements will be interesting in the
B-factories.
Even though those branching fractions are not measured directly,
ALEPH estimated
${\cal B}({\bar{B}}^0\rightarrow D^{+} \tau^- {\bar{\nu}})=(0.69\pm 0.14)\%$ and
${\cal B}({\bar{B}}^0\rightarrow D^{*+} \tau^- {\bar{\nu}})=(2.06\pm 0.41)\%$
\cite{aleph1}
from their experimental result
${\cal B}(b\rightarrow \tau^- {\bar{\nu}}X)=(2.75\pm 0.30\pm 0.37)\%$
\cite{aleph2} of the inclusive decays
by assuming that one-fourth of $b\rightarrow \tau^- {\bar{\nu}}X$ involve a
$D^{+}$ meson and the other three-fourths involve a $D^{*+}$ meson.
The above ALEPH estimations correspond to
${\cal{B}}({\bar{B}}^0\rightarrow D^{+} \tau^- {\bar{\nu}})/
{\cal{B}}({\bar{B}}^0\rightarrow D^{+} e^- {\bar{\nu}})=0.29\pm 0.08$
and
${\cal{B}}({\bar{B}}^0\rightarrow D^{*+} \tau^- {\bar{\nu}})/
{\cal{B}}({\bar{B}}^0\rightarrow D^{*+} e^- {\bar{\nu}})=0.37\pm 0.08$ .
ALEPH used these estimations for the background rejection in the measurement
of $|V_{cb}|$ from
${\bar{B}}^0\rightarrow D^{(*)+} l^- {\bar{\nu}}$ decays \cite{aleph1}.
OPAL also used those estimations in the measurement
of $|V_{cb}|$ from ${\bar{B}}^0\rightarrow D^{*+} \tau^- {\bar{\nu}}$ decays
\cite{opal}.
On the other hand, the result of L3 of the inclusive decays is given by
${\cal B}(b\rightarrow \tau^- {\bar{\nu}}X)=(1.7\pm 0.5\pm 1.1)\%$
\cite{l3}.
In this paper we calculate the branching fractions of the exclusive semileptonic
decays ${\bar{B}}^0\rightarrow D^{(*)+} \tau^- {\bar{\nu}}$ within the HQET
by using the experimentally measured form factors
without using models for the form factors.
Therefore, our calculation is model independent and does not contain any
theoretical ambiguities coming from form factor models.
We obtain
${\cal B}({\bar{B}}^0\rightarrow D^{+} \tau^- {\bar{\nu}})=(0.52\pm 0.07)\%$ and
${\cal B}({\bar{B}}^0\rightarrow D^{*+} \tau^- {\bar{\nu}})=(1.22\pm 0.06)\%$,
which correspond to
${\cal{B}}({\bar{B}}^0\rightarrow D^{+} \tau^- {\bar{\nu}})/
{\cal{B}}({\bar{B}}^0\rightarrow D^{+} e^- {\bar{\nu}})=
0.278^{+0.049}_{-0.035}$
and
${\cal{B}}({\bar{B}}^0\rightarrow D^{*+} \tau^- {\bar{\nu}})/
{\cal{B}}({\bar{B}}^0\rightarrow D^{*+} e^- {\bar{\nu}})=
0.256^{+0.014}_{-0.013}$ .
Our result for the second ratio is
much smaller than the corresponding ALEPH estimation,
and the ratio
${\cal B}({\bar{B}}^0\rightarrow D^{+} \tau^- {\bar{\nu}}) :
{\cal B}({\bar{B}}^0\rightarrow D^{*+} \tau^- {\bar{\nu}})$
is about $1 : 2.35$ instead of $1 : 3$.
The sum of our results of the two exclusive modes is also
much smaller than the ALEPH value of the inclusive decays,
and it is close to the mean value of the L3 result.
K\"orner and Schuler also studied the decays
${\bar{B}}^0\rightarrow D^{(*)+} \tau^- {\bar{\nu}}$ by using the form factors
from the free quark
decay model and the spectator quark model \cite{koerner}.
In the present paper, we use the recent experimental results \cite{cleo95, cleo97}
for the form factors, so our calculation is model independent.

{}From Lorentz invariance one finds the decomposition of the
hadronic matrix element
in terms of hadronic form factors:
\begin{eqnarray}
& &<D^+(p)|J_\mu |{\bar{B}}^0(P)>
\ =\ (P+p)_\mu f_+(q^2) + (P-p)_\mu f_-(q^2)
\nonumber\\
&=&
\Bigl( (P+p)_\mu
-{M^2-m^2\over q^2}q_\mu \Bigr) \, F_1(q^2)
+{M^2-m^2\over q^2}\, q_\mu \, F_0(q^2),
\label{a1}
\end{eqnarray}
where $J_\mu = {\bar{c}}\gamma_\mu (1-\gamma_5) b$.
We use the following notations:
$M=m_B$ represents the initial meson mass,
$m=m_{D^{(*)}}$ the final meson mass,
$m_l$ the lepton mass,
$P=p_B$, $p=p_{D^{(*)}}$,
and $q_\mu =(P-p)_\mu$.
The form factors $F_1(q^2)$ and $F_0(q^2)$
correspond to $1^-$ and $0^+$ exchanges, respectively.
At $q^2=0$ we have the constraint
$F_1(0)=F_0(0)$,
since the hadronic matrix element in (\ref{a1}) is nonsingular
at this kinematic point.

The differential decay rate is given by
\begin{equation}
d\Gamma ={1\over (2\pi)^3}{1\over 32M^3}|{\cal M}|^2\, dq^2\, dt,\qquad
{\cal M}={G_F\over {\sqrt{2}}}V_{cb}l^{\mu}h_{\mu},
\label{n1}
\end{equation}
where $l^\mu ={\bar{u_l}}\gamma^\mu (1-\gamma_5)v_\nu$ and
$h_\mu =<D^+(p)|J_\mu |{\bar{B}}^0(P)>$.
We use the notations: $k=p_{l^-}$, ${\bar{k}}=p_{\bar{\nu}}$,
$q^2=(P-p)^2=(k+{\bar{k}})^2$, $t=(P-{\bar{k}})^2=(p+k)^2$,
and $u=(P-k)^2=(p+{\bar{k}})^2$.
From (\ref{n1}) we have
\begin{equation}
{d\Gamma\over dq^2\, dt}={1\over (2\pi)^3}{1\over 32M^3}
{G_F^2\, |V_{cb}|^2\over 2}l_{\mu\nu}h^{\mu\nu},
\label{n3}
\end{equation}
where
\begin{eqnarray}
l_{\mu\nu}&=&4(k_\mu {\bar{k}}_\nu +{\bar{k}}_\mu k_\nu -g_{\mu\nu}k\cdot {\bar{k}}
-i{\varepsilon}_{\mu\nu\rho\sigma}k^\rho {\bar{k}}^\sigma ),
\label{n4}\\
h_{\mu\nu}&=&<D^+(p)|J_\mu |{\bar{B}}^0(P)>^*<D^+(p)|J_\mu |{\bar{B}}^0(P)>.
\label{n5}
\end{eqnarray}
{}From (\ref{a1}), (\ref{n4}) and (\ref{n5}),
after some calculations we get
\begin{eqnarray}
{1\over 8}l_{\mu\nu}h^{\mu\nu}&=&
|f_+(q^2)|^2\Bigl( -t^2+t(M^2+m^2-q^2)-M^2m^2+m_l^2(t+{1\over 4}(q^2-m_l^2))\Bigr)
\nonumber\\
&&+{\rm Re}\Bigl( f_+(q^2)f_-(q^2)\Bigr) {1\over 2}m_l^2(2t+q^2-2m^2-m_l^2)
\nonumber\\
&&+|f_-(q^2)|^2{1\over 2}m_l^2{1\over 2}(q^2-m_l^2).
\label{n6}
\end{eqnarray}
The allowed range of the variable $t$ is given by
\begin{equation}
ME-{m_l^2\over q^2}(ME-m^2)-MK(1-{m_l^2\over q^2})\le t \le
ME-{m_l^2\over q^2}(ME-m^2)+MK(1-{m_l^2\over q^2}),
\label{n7}
\end{equation}
where the energy $E$ and the magnitude of three momentum $K$ of $D^{(*)+}$ meson
in the ${\bar{B}}^0$ meson rest frame are given by
\begin{equation}
E(q^2)={1\over 2M}(M^2+m^2-q^2),\qquad
K(q^2)={1\over 2M}((M^2+m^2-q^2)^2-4M^2m^2)^{1 / 2}.
\label{n8}
\end{equation}

After the $t$ integration of (\ref{n3}) over the allowed range given
in (\ref{n7}),
the $q^2$ distribution of the decay rate is given by
\begin{eqnarray}
&&{d\Gamma ({\bar{B}}^0\rightarrow D^+ l^- {\bar{\nu}})\over dq^2}=
{G_F^2\over 24\pi^3}\, |V_{cb}|^2\, K(q^2)\,
(1-{m_l^2\over q^2})^2\times
\label{aa3}\\
&&
[\, (K(q^2))^2\, (1+{1\over 2}{m_l^2\over q^2})\, |F_1(q^2)|^2
\, +\, M^2\, (1-{m^2\over M^2})^2\, {3\over 8}\, {m_l^2\over q^2}
|F_0(q^2)|^2\, ]\, ,
\nonumber
\end{eqnarray}
where the allowed range of $q^2$ is given by
\begin{equation}
m_l^2\le q^2\le (M-m)^2.
\label{aa3a}
\end{equation}
The formula (\ref{aa3}) was also given
by Khodjamirian et al. \cite{Khodjamirian}.
{}For $m_l=0$, (\ref{aa3}) is reduced to the
well-known formula
$d\Gamma / dq^2
=(G_F^2 / (24\pi^3))\, |V_{cb}|^2\,
(K(q^2))^3\,
|F_1(q^2)|^2$
and
$0\le q^2\le (M-m)^2$.

{}From the HQET, we have \cite{iw}
\begin{eqnarray}
F_1(q^2)&=&V(q^2)\ =\ A_0(q^2)\ =\ A_2(q^2)\ =\
{M+m\over 2{\sqrt{Mm}}}\,\, {\cal{F}} (y),
\label{c1}\\
F_0(q^2)&=&A_1(q^2)\ =\
{2{\sqrt{Mm}}\over M+m}\,\, {y+1\over 2}\,\, {\cal{F}} (y),
\nonumber
\end{eqnarray}
where $y=(M^2+m^2-q^2)/(2Mm)$.
We use the experimentally measured results
\cite{cleo95, cleo97}
\begin{eqnarray}
{\cal{F}} (y)&=&{\cal{F}} (1)[1-\rho^2 (y-1)]
\label{cleo1}\\
{\rm with}&&
\left\{ \begin{array}{lll}
\rho^2_{D}=0.59\pm 0.25\, ,\
&|V_{cb}|{\cal{F}}_{D} (1)\times 10^2=3.37
&\mbox{for ${\cal{F}}_{D} (y)$}\\[1ex]
\rho^2_{D^*}=0.84\pm 0.15\, ,\
&|V_{cb}|{\cal{F}}_{D^*} (1)\times 10^2=3.51
&\mbox{for ${\cal{F}}_{D^*} (y)\,\, ,$}
\end{array}\right. 
\nonumber
\end{eqnarray}
where we used the mean values of $|V_{cb}|{\cal{F}}_{D^{(*)}} (1)$
since the ratios of branching fractions are independent of these normalizations.

By using the form factors in (\ref{c1}) with
${\cal{F}}_{D} (y)$ in (\ref{cleo1}),
we obtain from (\ref{aa3})
the spectra presented in Fig. 1,
where we find that the spectrum for muon drops down near $q^2=0$.
(The spectrum for electron also drops down near the very end of $q^2=0$.)
In Fig. 2 we can see this nature in more detail.
At $q^2=m^2_{\pi^-}=0.019\ {\rm GeV}^2$, $d\Gamma / dq^2$ for muon has
substantially smaller value than that for electron, and then the formula
$[\Gamma ({\bar{B}}^0\rightarrow D^+\pi^-)]/
[d\Gamma ({\bar{B}}^0\rightarrow D^+ l^- {\bar{\nu}}) / dq^2|_{q^2=m^2_{\pi^-}}]
=6\pi f_{\pi}^2|a_1|^2|V_{ud}|^2X_{\pi}$
which is used
for the test of factorization holds for the electron spectrum,
but not for the muon spectrum
if we look at the $d\Gamma / dq^2$ spectrum in full detail.
We present the obtained
branching fractions
and their ratio in Table 1, which gives
${\cal{B}}({\bar{B}}^0\rightarrow D^+ \tau^- {\bar{\nu}})/
{\cal{B}}({\bar{B}}^0\rightarrow D^+ e^- {\bar{\nu}})=0.278^{+0.049}_{-0.035}$ .
Since the HQET provides good informations about
the heavy to heavy form factors,
our results for the exclusive $B$ to $D$ semileptonic decays are reliable
without theoretical model dependence.

{}From Lorentz invariance one finds the decomposition of the
hadronic matrix element
in terms of hadronic form factors:
\begin{eqnarray}
& &<D^{*+}(p)|J_\mu |{\bar{B}}^0(P)>
\nonumber\\
&=&\varepsilon^{*\nu}(p)\Bigl( (M+m)g_{\mu\nu }A_1(q^2)
-2{P_\mu P_\nu \over M+m}A_2(q^2)
+{q_\mu P_\nu \over M+m}A_3(q^2)
\nonumber\\
&&+i\varepsilon_{\mu\nu\rho\sigma}{P^\rho p^\sigma \over M+m}V(q^2)\Bigr) ,
\label{b1}
\end{eqnarray}
where $\varepsilon_{0123}=1$ and
\begin{equation}
2mA_0(q^2)=(M+m)A_1(q^2)-{M^2-m^2+q^2\over M+m}A_2(q^2)
+{q^2\over M+m}A_3(q^2).
\label{b2}
\end{equation}
The form factors $V(q^2)$, $A_1(q^2)$, $A_2(q^2)$ and $A_0(q^2)$
correspond to $1^-$, $1^+$, $1^+$ and $0^-$ exchanges, respectively.
At $q^2=0$ we have the constraint
$2mA_0(0)=(M+m)A_1(0)-(M-m)A_2(0)$,
since the hadronic matrix element in (\ref{b1}) is nonsingular
at this kinematic point.

After a rather lengthy calculation similar to the procedure from (\ref{n1}) to
(\ref{aa3}) for ${\bar{B}}^0\rightarrow D^{+} l^- {\bar{\nu}}$ decays,
we obtain the $q^2$ distribution of the decay rate for
${\bar{B}}^0\rightarrow D^{*+} l^- {\bar{\nu}}$ given by
\begin{eqnarray}
&&{d\Gamma ({\bar{B}}^0\rightarrow D^{*+} l^- {\bar{\nu}})\over dq^2}=
{G_F^2\over 32\pi^3}\, |V_{cb}|^2\, {1\over M^2} K(q^2)\,
(1-{m_l^2\over q^2})^2\times
\label{b4}\\
&&\{ |A_1(q^2)|^2{(M+m)^2\over m^2}[{1\over 3}(MK)^2(1-{m_l^2\over q^2})
+q^2m^2+(MK)^2{m_l^2\over q^2}+{1\over 2}m^2m_l^2]
\nonumber\\
&&\ +{\rm Re}(A_1(q^2)A_2^*(q^2))[
-{M^2-m^2-q^2\over m^2}
[{2\over 3}(MK)^2(1-{m_l^2\over q^2})
+2(MK)^2{m_l^2\over q^2}
\nonumber\\
&&\qquad\qquad\qquad\qquad +{1\over 2}(M^2+m^2-q^2)m_l^2]
+(M^2-m^2+q^2)m_l^2]
\nonumber\\
&&\ +|A_2(q^2)|^2{1\over (M+m)^2m^2}(MK)^2[{4\over 3}(MK)^2(1-{m_l^2\over q^2})
+4(MK)^2{m_l^2\over q^2}+2M^2m_l^2]
\nonumber\\
&&\ +|V(q^2)|^2{q^2\over (M+m)^2}[{8\over 3}(MK)^2(1-{m_l^2\over q^2})
+4(MK)^2{m_l^2\over q^2}]
\nonumber\\
&&\ +|A_3(q^2)|^2{q^2\over (M+m)^2m^2}{1\over 2}(MK)^2m_l^2
\nonumber\\
&&\ -{\rm Re}(A_3(q^2)A_2^*(q^2)){1\over (M+m)^2m^2}(M^2-m^2+q^2)(MK)^2m_l^2
\nonumber\\
&&\ +{\rm Re}(A_3(q^2)A_1^*(q^2)){1\over m^2}(MK)^2m_l^2\} .
\nonumber
\end{eqnarray}
When $m_l=0$ is taken in the formula (\ref{b4}), it agrees with the formula
for $m_l=0$
given in Ref. \cite{ukqcd},
and when we use the relations (\ref{c1}) of the HQET with $m_l=0$,
(\ref{b4}) reduces to
the well-known formula
$d\Gamma / dq^2 =
(G_F^2 / (48\pi^3))\, |V_{cb}|^2\, m^3\, (M-m)^2\, {\sqrt{y^2-1}}\,
(y+1)^2\,
\{ 1+(4y / (y+1))\, ((1-2yr+r^2) / (1-r)^2) \} \, 
({\cal{F}}_{D^*} (y))^2$,
where $r=m/M$.

By using the form factors given
in (\ref{c1}) with
${\cal{F}}_{D^*} (y)$ in (\ref{cleo1}),
we obtain from (\ref{b4}) the spectra
presented in Fig. 3.
We present the obtained branching fractions
and their ratio in Table 1, where
we find that
${\cal{B}}({\bar{B}}^0\rightarrow D^{*+} \tau^- {\bar{\nu}})/
{\cal{B}}({\bar{B}}^0\rightarrow D^{*+} e^- {\bar{\nu}})=
0.256^{+0.014}_{-0.013}$
which is much smaller than the ALEPH estimation $0.37\pm 0.08$ \cite{aleph1}.

%We derived the formulas for the $q^2$ distributions of the decay rate for
%${\bar{B}}^0\rightarrow D^{*+} \tau^- {\bar{\nu}}$ decays.

\pagebreak

%\vspace*{0.5cm}

\noindent
{\em Acknowledgements} \\
\noindent
This work was supported
by Non-Directed-Research-Fund,
Korea Research Foundation 1997,
by the Basic Science Research Institute Program,
Ministry of Education, Project No. BSRI-97-2414,
by the Korea Science and Engineering Foundation,
Grant 985-0200-002-2,
and by the Research Fund of
Kangnung National University 1997.
\\

\vspace*{0.7cm}

%\pagebreak

\pagebreak

\vspace*{1.0cm}

\begin{table}[h]
\vspace*{1.5cm}
\hspace*{-1.8cm}
%\begin{center}
\begin{tabular}{|c|c|c|c|c|}   \hline
&${\cal{B}}(D^{(*)+} e^-)\times 10^{2}$
&${\cal{B}}(D^{(*)+} \mu^-)\times 10^{2}$
&${\cal{B}}(D^{(*)+} \tau^-)\times 10^{2}$
&${\cal{B}}(e^-) :
{\cal{B}}(\mu^-) :
{\cal{B}}(\tau^-)$
\\   \hline
$\ \ D^{+}\ \ $&$1.85^{-0.47}_{+0.56}$&$1.84^{-0.47}_{+0.56}$
&$0.52^{-0.07}_{+0.07}$
&1\, :\, $0.996^{+0.001}_{-0.001}$\, :\, $0.278^{+0.049}_{-0.035}$\\
$\ \ D^{*+}\ \ $&$4.76^{-0.47}_{+0.50}$&$4.74^{-0.46}_{+0.50}$
&$1.22^{-0.06}_{+0.06}$
&1\, :\, $0.996^{+0.002}_{-0.000}$\, :\, $0.256^{+0.014}_{-0.013}$\\
\hline
\end{tabular}
%\end{center}
\caption{The obtained branching fractions and their ratio
for ${\bar{B}}^0\rightarrow D^{(*)+} l^- {\bar{\nu}}$.}
\end{table}

%\pagebreak

\vspace*{5.0cm}

\noindent
{\large\bf
Figure Captions}\\

\noindent
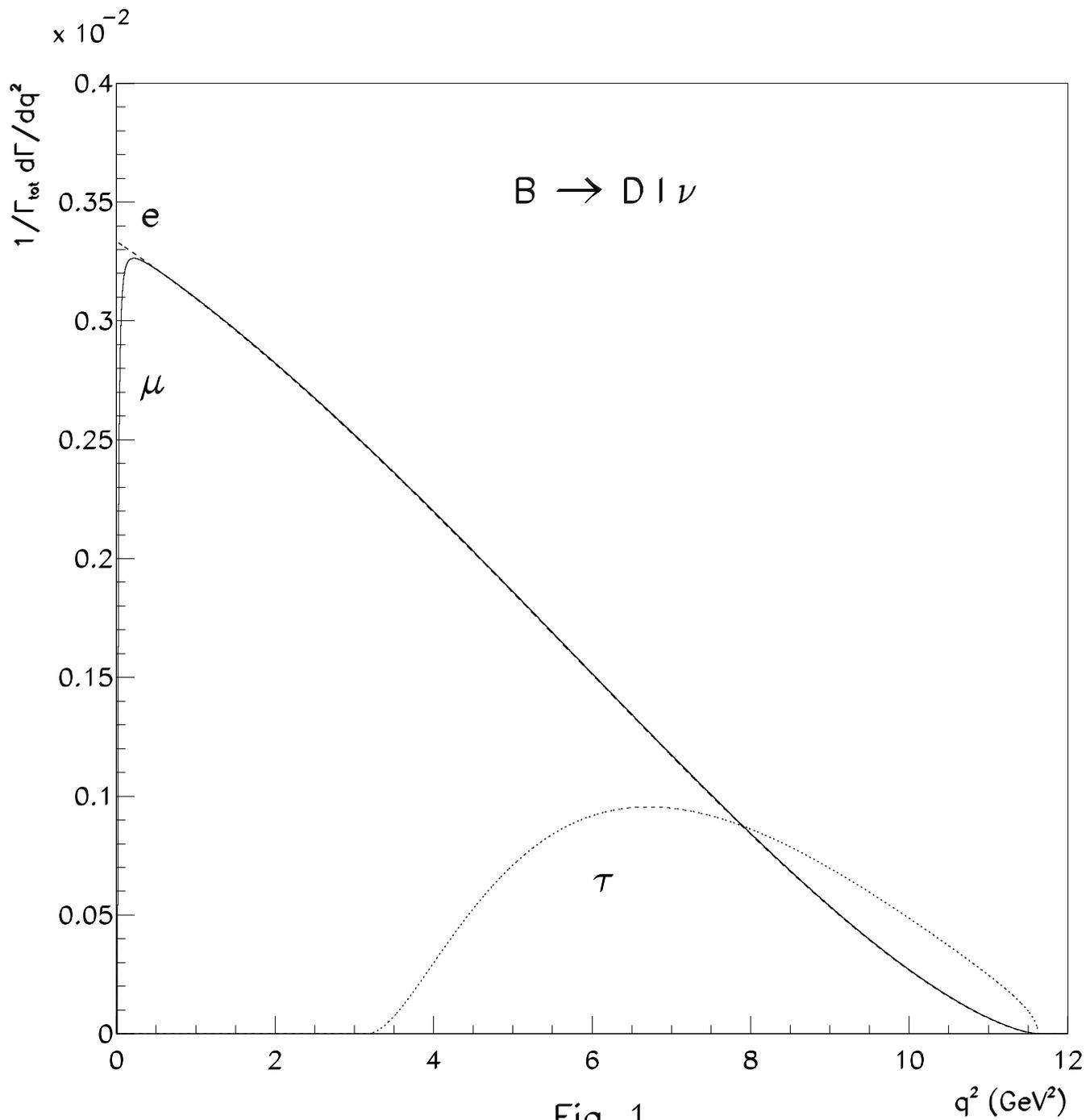
Fig. 1. $(1/\Gamma_{tot})(d\Gamma/dq^2)$ of
${\bar{B}}^0\rightarrow D^{+} l^- {\bar{\nu}}$.\\

\noindent
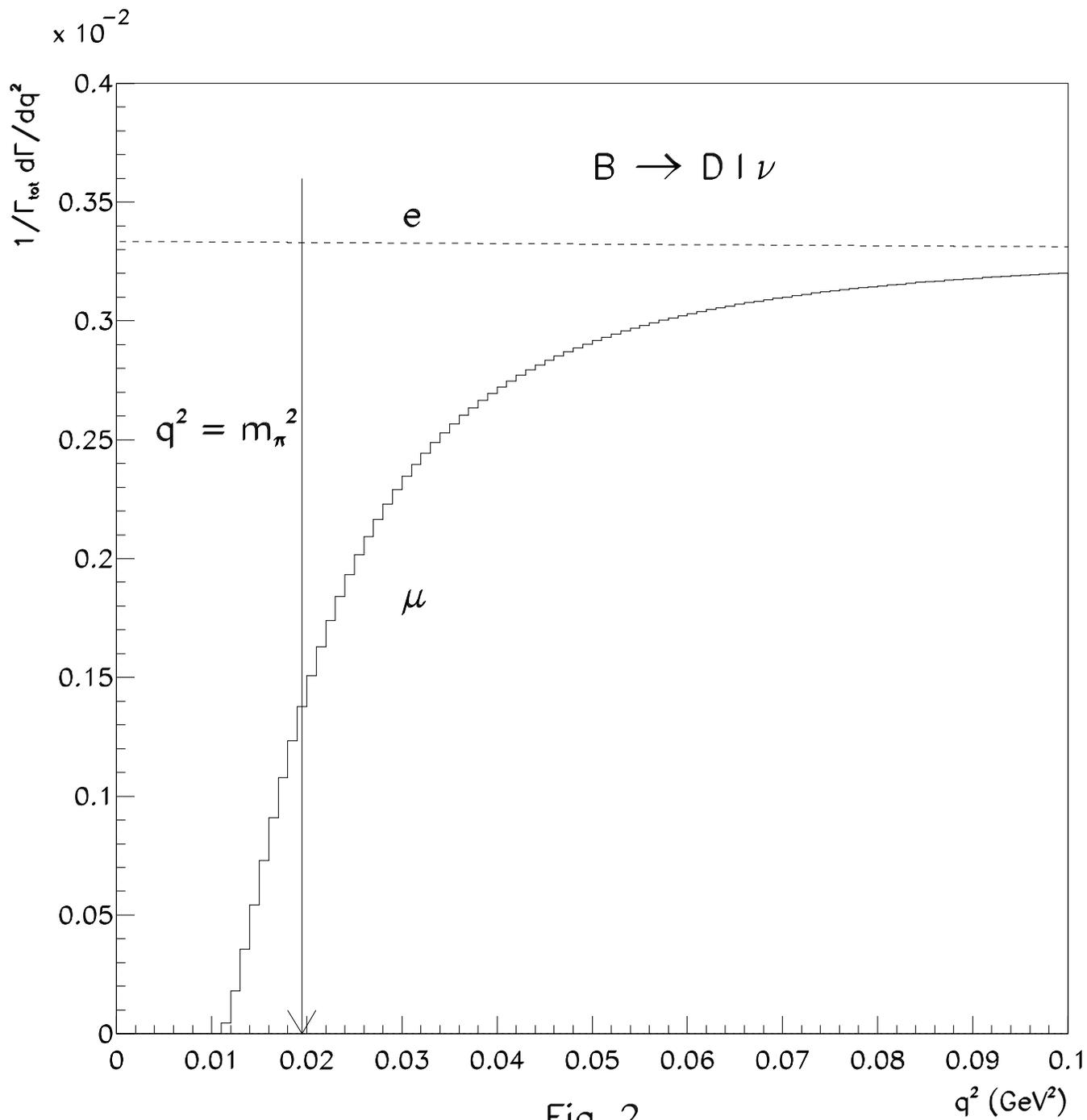
Fig. 2. $(1/\Gamma_{tot})(d\Gamma/dq^2)$ of
${\bar{B}}^0\rightarrow D^{+} l^- {\bar{\nu}}$
for $0\le q^2 \le 0.1\ {\rm GeV}^2$.\\

\noindent
Fig. 3. $(1/\Gamma_{tot})(d\Gamma/dq^2)$ of
${\bar{B}}^0\rightarrow D^{*+} l^- {\bar{\nu}}$.\\

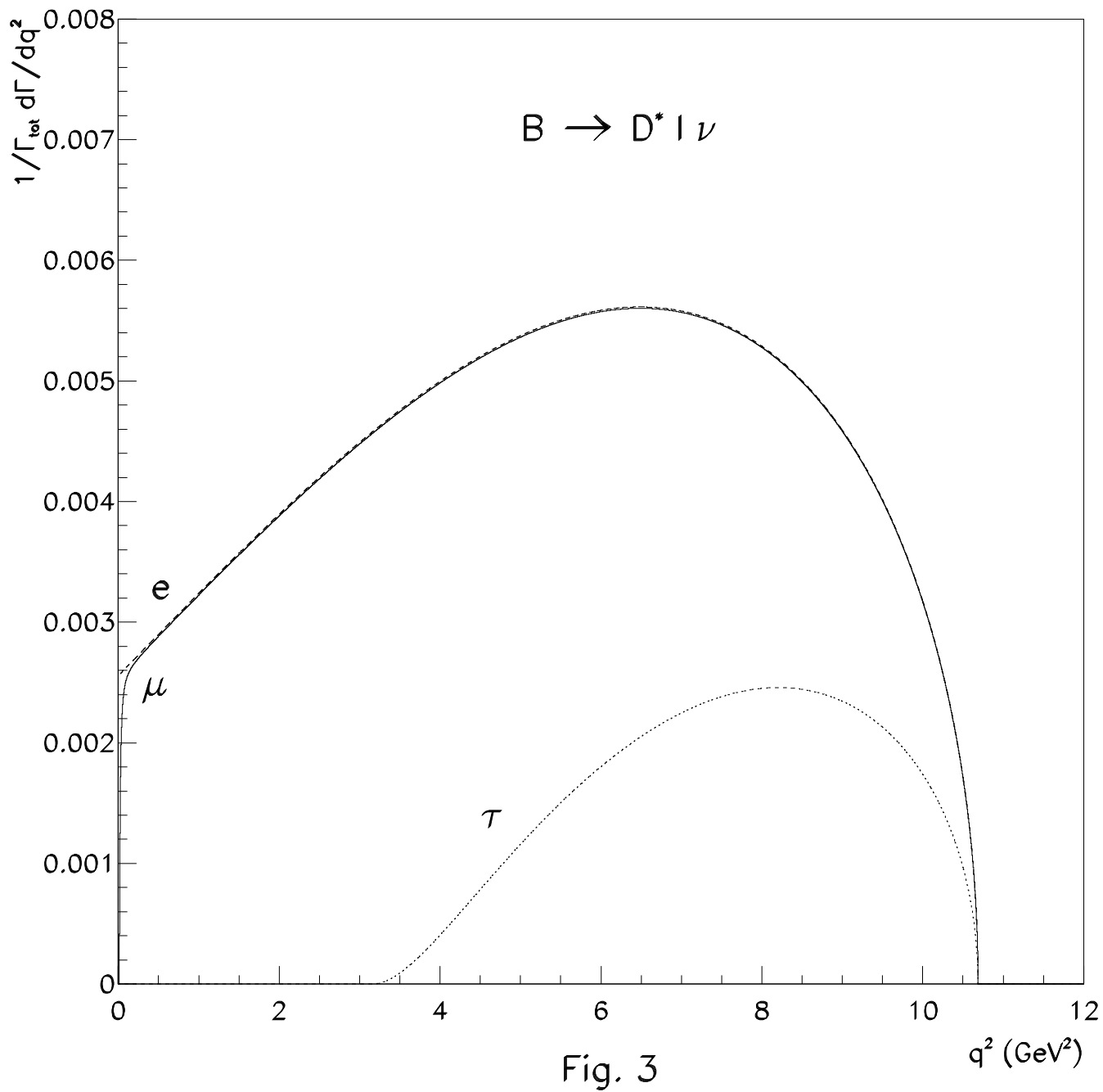
\end{document}